\documentclass[useAMS,usenatbib]{mn2e}

\usepackage[pdftex]{graphicx}

\usepackage{psfrag}
\usepackage{color}

\def\solmasp{$\mathrm{M_\odot}$}

\def\be{\begin{equation}}
\def\ee{\end{equation}}
\def\log{\mathrm{log}}

\def\thetatol{$\theta_{\rm tol}$}
\def\treecol{{\em TreeCol}}

\def\aap{{ A\&A}}

\def\apjl{{ApJL}}

\def\mnras{{MNRAS}}

\def\nat{Nature}

\def\apjs{{ApJS}}

\title[Estimating column densities in astrophysical simulations]
{{\em TreeCol}: a novel approach to estimating column densities in astrophysical simulations}
\author[Clark, Glover \& Klessen]
{Paul C. Clark\thanks{E-mail: p.clark@uni-heidelberg.de}, Simon C.O. Glover and Ralf S. Klessen
\\ Institut f\"ur theoretische Astrophysik, Zentrum f\"ur Astronomie  der Universit\"at Heidelberg,
Albert-Ueberle-Stra\ss e 2,  \\ 69120 Heidelberg, Germany
}

\date{\today}

\begin{document}
\maketitle

%
%

\begin{abstract}
We present {\em TreeCol}, a new and efficient tree-based scheme to calculate column densities in numerical simulations. Knowing the column density in any direction at any location in space is a prerequisite for modeling the propagation of radiation  through the computational domain. {\em TreeCol} therefore forms the basis for a fast, approximate method for modelling the attenuation of radiation within large numerical simulations. It constructs a {\em HEALPix} sphere at any desired location and accumulates the column density by walking the tree and by adding up the contributions from all tree nodes whose line of sight contributes to the pixel under consideration. In particular when combined with widely-used tree-based gravity solvers the new scheme requires little additional computational cost. In a simulation with $N$ resolution elements, the computational cost of {\em TreeCol} scales as $N \log N$, instead of the $N^{5/3}$ scaling of most other radiative transfer schemes. {\em TreeCol} is naturally adaptable to arbitrary density distributions and is easy to implement and to parallelize, particularly if a tree structure is already in place for calculating the gravitational forces. We describe our new method and its implementation into the SPH code Gadget~2. We discuss its accuracy and performance characteristics for the examples of a spherical protostellar core and for the turbulent interstellar medium. We find that the column density estimates provided by {\em TreeCol} are on average accurate to better than 10 percent. In another application, we compute the dust temperatures for solar neighborhood conditions and compare with the result of a full-fledged Monte Carlo radiation-transfer calculation. We find that both methods give very similar answers. We conclude that {\em TreeCol} provides a fast, easy to use, and sufficiently accurate method of calculating column densities that comes with little additional computational cost when combined with an existing tree-based gravity solver.
\end{abstract}

%
%

\section{Introduction}
The penetration of radiation into an optically thick distribution of gas is a feature of many astrophysical systems, ranging from scales as small as those of circumstellar disks  to those as large as the damped Lyman-$\alpha$ absorbers observed along the sightlines to many quasars. Numerical modelling of the propagation of radiation through the gas can greatly aid our efforts to understand the astrophysics of these systems, but frequently proves to be computationally challenging, owing to the high dimensionality of the problem. In the common case in which we have no useful spatial symmetries to exploit and wish to solve for the properties of the radiation field within $N_{\nu}$ different frequency bins, the computational cost of determining the full spatial and angular distribution of the radiation field is of order $N^{5/3} \times N_{\nu}$, where $N$ is the number of resolution elements (e.g.\ grid cells in an Eulerian simulation, or particles in a smoothed particle hydrodynamics [SPH] model), and where we have assumed that the desired angular resolution is comparable to the spatial resolution. For static problems, where the gas distribution is fixed and we need only to solve for the properties of the radiation field at a single point in time, it is currently possible to solve the full radiative transfer problem numerically even for relatively large values of $N$ \citep[see e.g.][who post-process the results of an SPH simulation with $N = 3.5 \times 10^{6}$]{rundle10}.
However, if one is interested in dynamical problems, where the gas distribution is not fixed and the gas and radiation significantly influence one another, then the cost of solving for the radiation field after every single hydrodynamical timestep can easily become prohibitively large (for a detailed discussion, see \citealt{kkh09}).

For this reason, it is useful to look for simpler, more approximate techniques for treating the radiation that have a much lower computational cost, and that can therefore be used within hydrodynamical simulations without rendering these simulations overly expensive. One common simplification that nevertheless has a reasonably broad range of applicability is to ignore the re-emission of incident radiation within the gas. Making this simplification means that rather than solving the full transfer equation,
\begin{equation}
\frac{\partial I_{\nu}}{\partial s} = \eta_{\nu} - \chi_{\nu} I_{\nu} \label{full_rt}
\end{equation}
along multiple rays through the gas, where $I_{\nu}$ is the specific intensity at a frequency $\nu$, $\eta_{\nu}$ and $\chi_{\nu}$ are the emissivity and opacity at the same frequency, and $s$ is the path length along the ray, one can instead solve the simpler equation,
\begin{equation}
\frac{\partial I_{\nu}}{\partial s} = - \chi_{\nu} I_{\nu}. \label{simple_rt}
\end{equation}
Equation~\ref{full_rt} has the formal solution
\begin{equation}
I_{\nu} = I_{\nu, 0} e^{-\tau_{\nu}} + \int_{0}^{\tau_{\nu}} S_{\nu} e^{-\tau_{\nu}^{\prime}} {\rm d}\tau_{\nu}^{\prime},   \label{formal_full}
\end{equation}
where $I_{\nu, 0}$ is the specific intensity of the radiation field at the start of the ray (e.g.\ at the edge of a gas cloud), $S_{\nu} \equiv 
\eta_{\nu} / \chi_{\nu}$ is the source function, and $\tau_{\nu}$ is the optical depth along the ray. If $S_{\nu} \ll I_{\nu, 0}$ and the
optical depth is not too large, then it is reasonable to neglect the integral term, in which case we can write $I_{\nu}$ as
\begin{equation}
I_{\nu} = I_{\nu, 0} e^{-\tau_{\nu}},
\end{equation}
which is the formal solution to Equation~\ref{simple_rt}. By making this approximation, we therefore reduce the problem to one of determining optical depths along a large number of rays. Often, this problem can then be further reduced to one of determining the column density of some absorber (e.g.\ dust) along each ray. 

Unfortunately, although these simplifications make the problem easier to handle numerically, they do not go far enough, as the most obvious technique for calculating the column densities -- integrating along each ray -- still has a computational cost that scales as $N^{5/3}$ and hence is impractical in large simulations. This motivates one to look for computationally cheaper methods for determining the angular distribution of column densities seen by each resolution element within a large numerical simulation. 

In this paper, we introduce a computationally cheap and acceptably accurate method for computing these column densities, suitable for use within simulations of self-gravitating gas that utilize a tree-based solver for calculating gravitational forces. Our method, which we dub \treecol, makes use of the large amount of information on the density distribution of the gas that is already stored within the tree structure to accelerate the calculation of the required column density distributions.

In the next section, we give a description of how our algorithm works, starting with a overview of how tree-based gravity solvers work in Section \ref{overview}, and then showing how it is possible to implement the \treecol~method in Section \ref{implement}. We then present two stringent tests of the algorithm in Section \ref{tests}, both of which are typical of the conditions found in contemporary simulations of star formation. We discuss some of the potential applications of the \treecol~method in Section \ref{applications}. In Section \ref{performance}, we give an overview of the computational efficiency of this scheme, and we summarise this paper in Section \ref{summary}.

\section{TreeCol}

\begin{figure}
\centerline{
\includegraphics[width=2.6in]{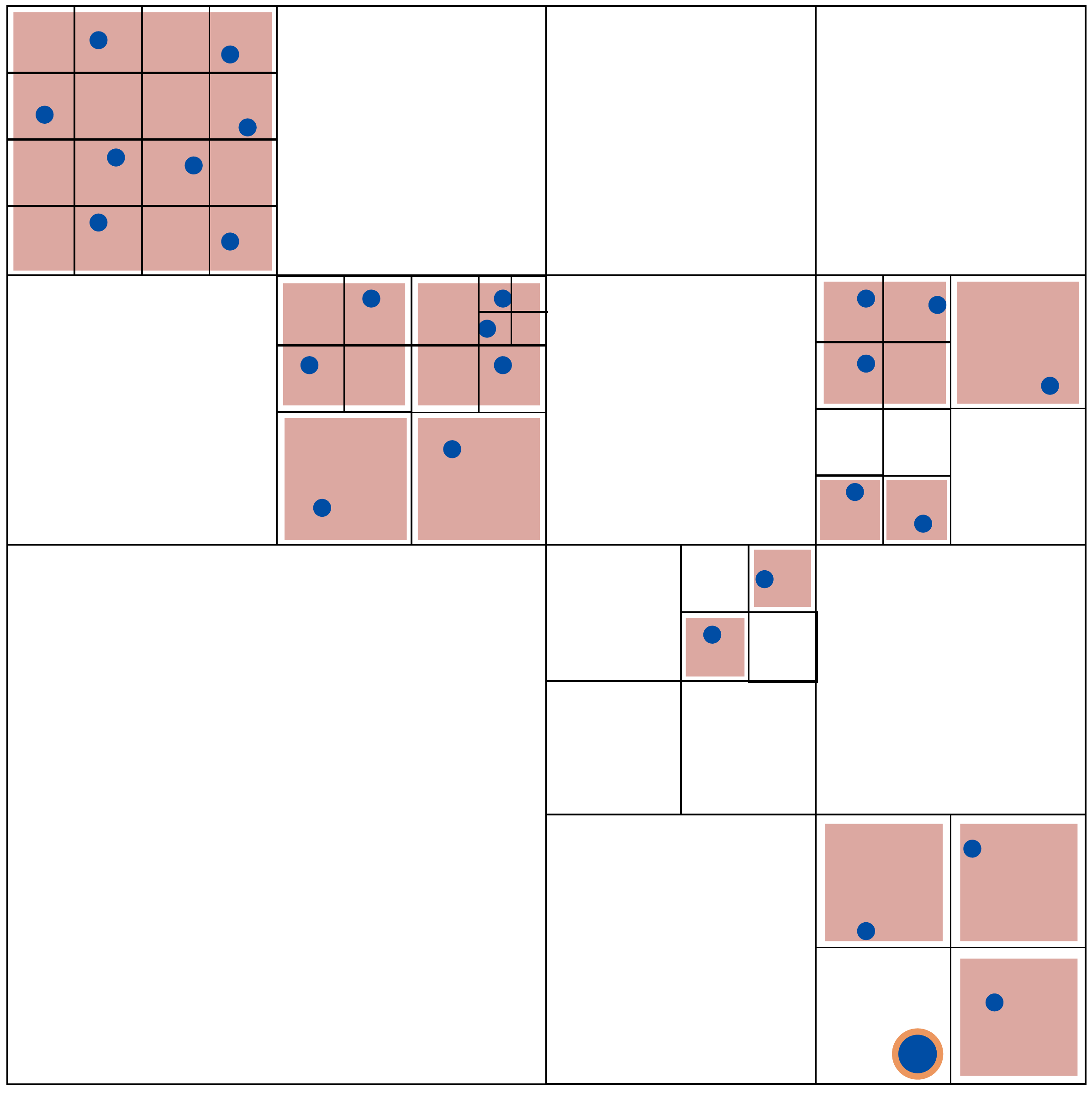}
}
\caption{\label{fig:tree}Schematic diagram showing how the tree is constructed and used for the gravitational force calculation. A 3D oct-tree splits each parent node into
eight daughter nodes, but in this 2D representation, we show only four of these nodes. The black lines show the boundaries of the tree nodes that would be constructed for the given ensemble of particles, shown as blue dots. The regions shaded in red denote the nodes that would be used to calculate the gravitational force as seen by the large blue and orange particle at the bottom of the diagram. Note that in the case where the nodes being used contain only one particle (a `leaf' node), the position of the particle itself is used to calculate the gravitational force arising from that node.}
\end{figure}

\begin{figure}
\centerline{\includegraphics[width=3.in]{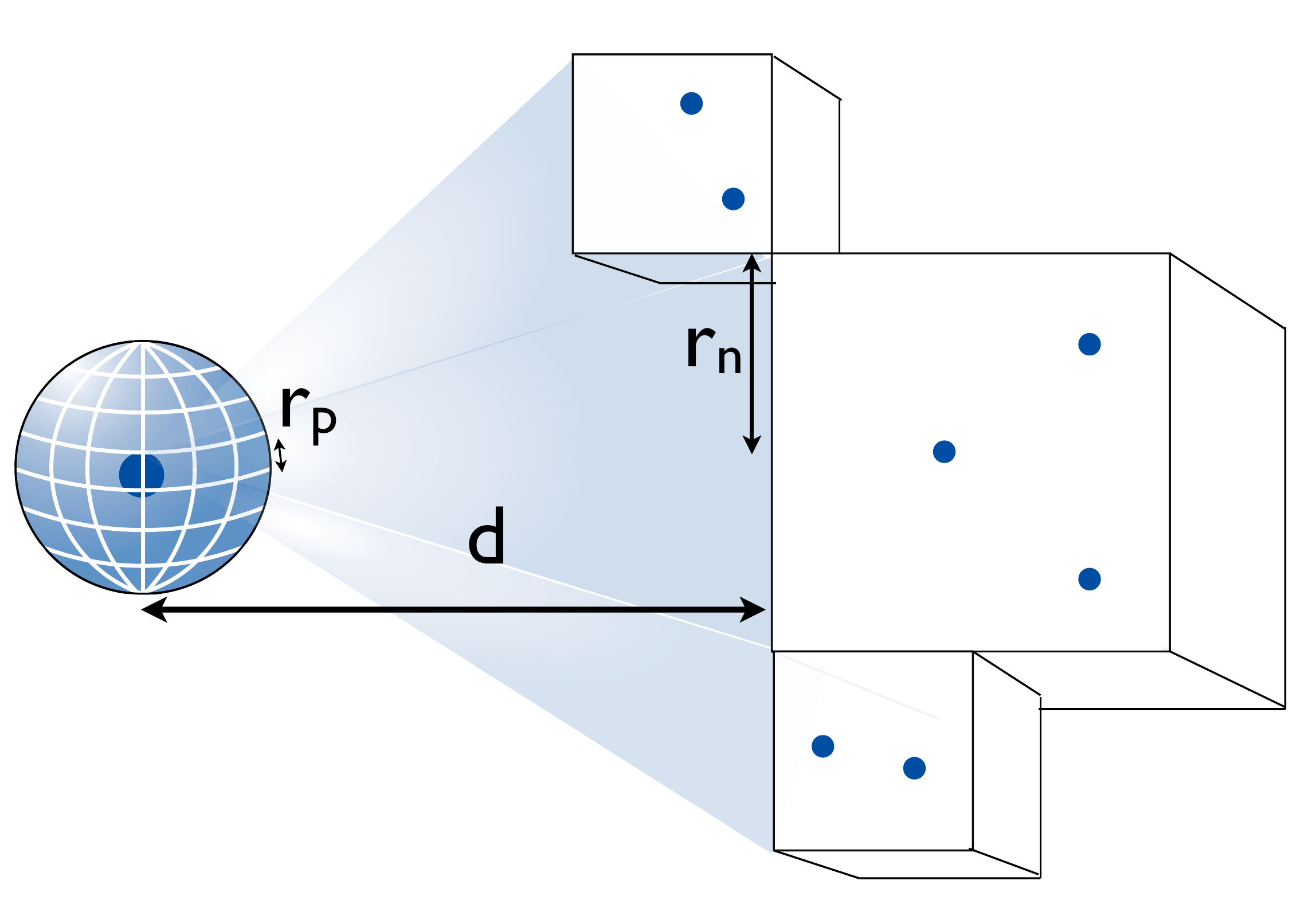}}
\caption{\label{fig:treecoldia} Schematic diagram illustrating the \treecol~concept. During the tree walk to obtain the gravitational forces, the projected column densities of the tree nodes (the boxes shown on the right) are mapped onto a spherical grid surrounding the particle for which the forces are being computed (the ``target'' particle,
shown on the left). The tree already stores all of the information necessary to compute the column density of each node, the position of the node in the plane of the sky
of the target particle, and the angular extent of the node. This information is used to compute the column density map at the same time that the tree is being walked to
calculate the gravitational forces. Provided that the tree is already employed for the gravity calculation, the information required to create the $4\pi$ steradian map of the column densities can be obtained for minimal computational cost.}
\end{figure}

\subsection{Basic idea behind \treecol}
\label{overview}
Tree-based gravity solvers (e.g. \citealt{bh86, bh89}) have long been a standard feature of $N$-body and smoothed particle hydrodynamics codes (e.g. \citealt{Benz88, VineSigurdsson98, springel01, Wadsley2004}).  More recently, their accuracy and speed has also seen them adopted in grid-based codes \citep{daleetal09}. In this paper, we describe a method whereby the information stored in the gravitational tree can be used to construct a $4\pi$ steradian map of the column density. By constructing this map at the same time as the
tree is being ``walked'' to determine the gravitational forces, we can minimize the amount of additional communication necessary between CPUs holding different portions of the tree. Since the structure of the tree, and how it is walked, will be important for our discussion, we will first give a brief overview of how a tree-based gravity solver works. For the purpose of this discussion, we consider a solver based on an oct-tree, as used in e.g.\ the Gadget SPH code \citep{springel05}, although we note that
solvers based on other tree structures, such as binary trees, do exist (e.g.\ the binary tree
employed by \citealt{Benz88}, which later found its way into other high profile studies, such as 
\citealt{Bonnell98} and \citealt{Bate03}).

A tree-based solver starts by constructing a tree, splitting the computational volume up into a series of nested boxes, or `nodes'. The `root' node is the largest in the hierarchy and contains all of the computational points in the simulation. This large `parent' node is then split up into eight smaller `daughter' nodes  as shown in Figure~\ref{fig:tree}. The daughter nodes are further refined (becoming parents themselves) until each tree node contains only one particle (illustrated  in Figure~\ref{fig:tree} by the blue dots). These smallest nodes at the very bottom of the hierarchy are typically termed `leaves'. At each point in the hierarchy, the tree stores the information about the contents of the parent node (including its position, mass and size) that will be needed during the gravitational force calculation. Once the construction of the tree is complete, each particle is located in a leaf node situated at the bottom of a nested hierarchy of other nodes. 

Once the tree is built, it can then be ``walked'' to get the gravitational forces. The idea behind the speed-up offered by the tree gravity solver over direct summation is very simple: any region of structured mass that is far away can be well approximated as a single, unstructured object, since the distances to each point in the structure are essentially the same.  Strictly, this is only true if the angular size of the structure is small, and so tree-codes tend to adopt an angle, rather than a distance, for testing whether or not
structures can be approximated. This angle is often referred to as the ``opening angle'' of the tree, and we will denote it hereafter as \thetatol.

To walk the tree to obtain the gravitational force on a given particle, the algorithm starts at the root node and opens it up, testing whether the daughter nodes subtend an angle of less than \thetatol. If the angle is smaller than \thetatol, the properties of the daughter nodes (mass, position, centre of mass) are used to calculate their contribution to the force. As such, any substructure within the daughter nodes is ignored, and the mass inside in the nodes is assumed to be uniformly distributed within their boundaries. If one or more of these nodes subtends an angle larger than \thetatol, the nodes are opened and the process is repeated on their daughter nodes, and so on, until nodes are found that appear smaller than \thetatol. To increase the accuracy of the force calculation, the nodes often store multipole moments that account for the fact that the node is not a point mass, but rather a distributed object that subtends some finite angle (e.g. see \citealt{bt87}). These moments are calculated during the tree construction, for all levels of the node hierarchy except the leaves, since these are either well approximated as point masses -- as is the case for a stellar $N$-body calculation -- or are SPH particles, which have their own prescription for how they are distributed in space \citep{bbp95}.   

The above method is sketched in Figure \ref{fig:tree}, which shows the tree structure in black, and the nodes, marked in red, that would be used to evaluate the gravitational force on the large blue particle with the orange highlight. In the cases where the nodes are leaves (containing only a single particle), the position of the particle itself is used. As the total number of force calculations can be substantially decreased in comparison to the number required when using direct summation, tree-based gravity solvers offer a considerable speed-up at the cost of a small diminution in accuracy. \citet{bh89} showed that for a distribution of $N$ self-gravitating particles, the computational cost of a tree-based solver scales as $N\,{\rm log}\, N$, compared to the $N^2$ scaling associated with direct summation. They also showed that the multipole moments allowed quite large opening angles, with \thetatol~values as large as 0.5 radians resulting in errors of less than a percent.  

Our \treecol ~method makes use of the fact that each node in the tree stores the necessary properties for constructing a column density map. The mass and size of the node can be used to calculate the column density of the node, and its position and apparent angular size allow us to determine the region on the sky that is covered by the node. Note also that column density, just like the total gravitational force, is a simple sum over the contributing material, meaning that it is independent of the order in which the contributions are gathered. Just as the tree allows us to construct a force for each particle, we can also sum up the column density contributions of the nodes to create a $4\pi$ steradian map of the column density during the tree-walk.

A schematic diagram of how this works is shown in Figure \ref{fig:treecoldia}. The target particle -- the one currently walking the tree, and for which the map is being created -- is shown as the large dark blue particle on the left. Around it we show the spherical grid onto which the column densities are to be mapped. We see that the tree nodes, shown on the right, subtend some angle $\theta$ (which is less than some adopted \thetatol), and cover different pixels on the spherical grid. During the tree walk, the \treecol ~method simply maps the projection of the nodes onto the pixels for the particle being walked. 

\begin{figure}
\centerline{\includegraphics[width=3.2in]{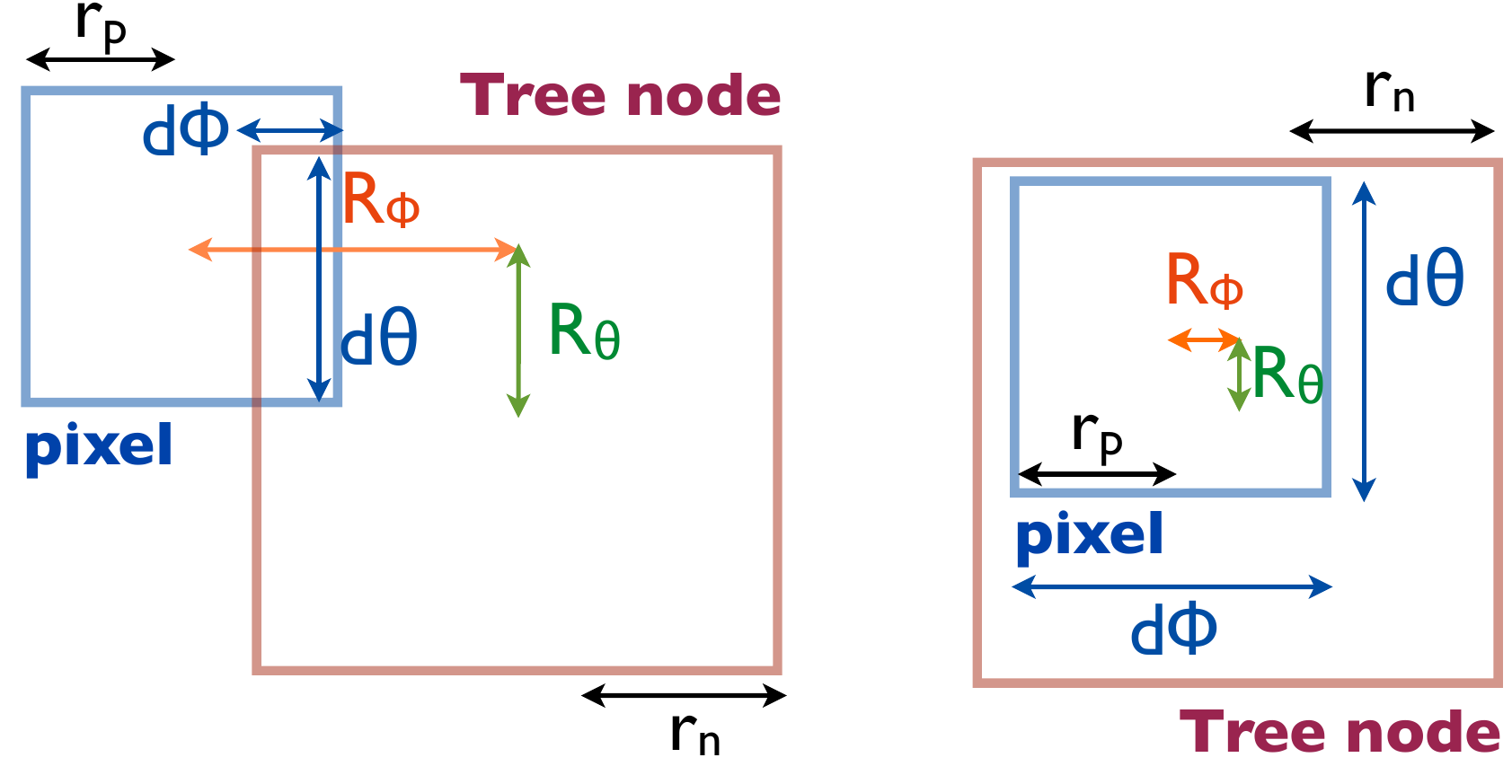}}
\caption{\label{fig:pixdia} Schematic showing the overlap between a pixel on the SPH particle's {\em HEALPix} sphere, and the tree node. The angular size of the pixels and nodes are denoted by $2r_{\rm p}$ and $2r_{\rm n}$, respectively, and the distances between their centres are given by the orthogonal angles $R_\theta$ and $R_\phi$. The diagram shows the case when the angle subtended by the tree node is greater than that of the pixels, and the two possible situations that can arise: a) the pixel and tree node only partially overlap, and b) the pixel is entirely covered by the tree node. In the former case, we work out the mass in the overlapping area, and convert it to a column density contribution by smearing it over the pixel's area. In the latter case, the pixel just obtains the full column density of the node. In the case where the angle subtended by the pixels is greater than the tree node (not shown here), then obviously the tree node can also become totally covered by the pixel. In this case, the full mass of the node is smeared out over the pixel's area to define the column density contribution. Full details of how the mapping is done in this implementation are given in Section \ref{implement}.}
\end{figure}

\begin{figure}
\centerline{\includegraphics[width=3.2in]{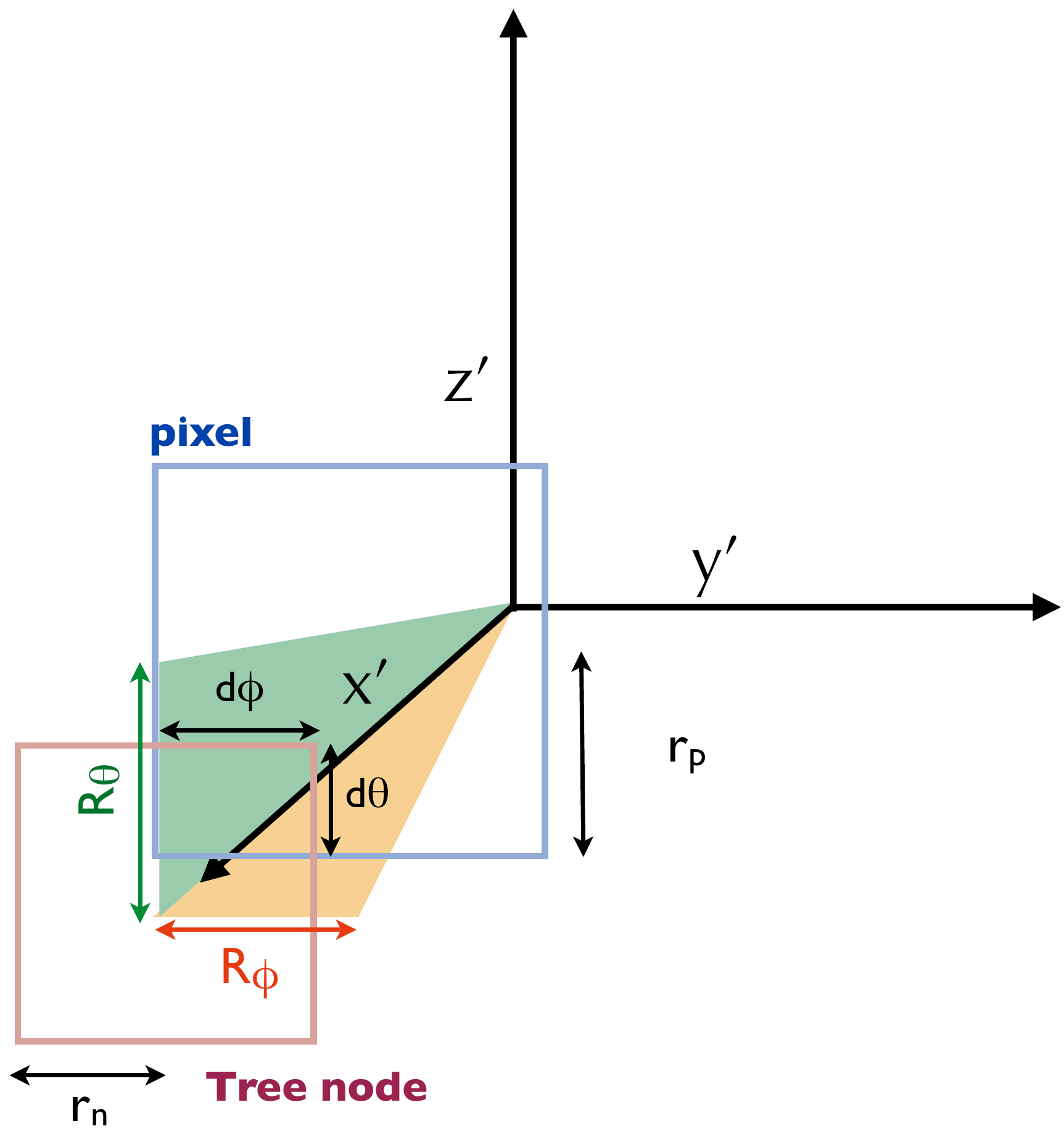}}
\caption{\label{fig:nodepro} Illustration showing how the nodes and pixels are assumed to interact in our implementation of the \treecol~algorithm. For each tree node, a new co-ordinate system is created, in which the node's position vector is the $x$-axis. The angular distance between the node centres, can then be described by two orthogonal angles, $R_\theta$ and $R_\phi$, which allows us to define an overlap area $d\phi \,d\theta$. Note that nodes and pixels have an area $(2 r_{\rm n})^2$ and $(2 r_{\rm p})^2$ respectively. Full details are given in Section \ref{implement}.}
\end{figure}

\subsection{A simple implementation of \treecol}
\label{implement}

The details of exactly how the nodes are mapped onto the grid depends on how accurate one needs the column density information to be. However,  it should be noted that the tree structure is only an approximate representation of the underlying gas structure: it distributes the mass in a somewhat larger volume than is actually the case, and as a result, sharp edges tend to be displaced to the boundary of node. As such, column densities from the tree will always be approximate, and so a highly accurate mapping of the node column density projections is computationally wasteful. In what follows, we will describe a simple implementation of \treecol~that is both reasonably accurate while at the same time requiring minimal computational cost.

Our mapping of the tree nodes to the pixels makes a number of assumptions regarding the shape and projection of the nodes and the pixels. These are:

\begin{itemize}
\item The tree nodes are always seen as squares in the sky, regardless of their actual orientation.
\item The nodes are assumed to overlap the pixels as shown in Figure \ref{fig:nodepro}, such that we can define the overlapping region based on simple orthogonal co-ordinates in the plane of the sky.
\item We use the {\em HEALPix}\footnote{http://healpix.jpl.nasa.gov/} algorithm \citep{healpix} to compute pixels that are equidistant on the sphere's surface and that have equal areas.
\end{itemize}

We show a schematic diagram of the way the nodes are assumed to overlap in Figure \ref{fig:pixdia}. The tree nodes are taken to be squares with side length $2r_{\rm n}$ and likewise, the pixels onto which the column densities mapped are assumed to be squares with side length $2r_{\rm p}$. As shown in the diagram, these dimensions are assumed to be equivalent to the angles subtended by the nodes and the pixels. Overlap requires that
\begin{equation}
R_{\theta} < r_{\rm p} + r_{\rm n}
\end{equation}
\noindent and
\begin{equation}
R_{\phi} < r_{\rm p} + r_{\rm n}.
\end{equation}
If this is the case, the lengths, $d\theta$ and $d\phi$ describing the overlapping area are then given by
\begin{equation}
d\theta = {\rm min}\{ (r_{\rm p} + r_{\rm n} - R_{\theta}),  2r_{\rm p}\}
\end{equation}
and
\begin{equation}
d\phi = {\rm min}\{ (r_{\rm p} + r_{\rm n} - R_{\phi}),  2r_{\rm p}\},
\end{equation}
when the pixels have a smaller angular extent than the nodes (i.e.\ $r_{\rm p} < r_{\rm n}$), or
\begin{equation}
d\theta = {\rm min}\{ (r_{\rm p} + r_{\rm n} - R_{\theta}),  2r_{\rm n}\}
\end{equation}
and
\begin{equation}
d\phi = {\rm min}\{ (r_{\rm p} + r_{\rm n} - R_{\rm \phi}),  2r_{\rm n}\},
\end{equation}
when the nodes have a smaller angular extent than the pixels (i.e.\ $r_{\rm n} < r_{\rm p}$). By taking the minimum of the expression $(r_{\rm p} + r_{\rm n} - R_{\theta,\phi})$ and either the node or pixel side length, we account for situations such as those shown in the right-hand panel in Figure~\ref{fig:pixdia}, in which either the pixel is totally covered by the node, or the node is totally contained within the pixel. We can then calculate the contribution of the node to the pixel's column density from
\begin{equation}
\label{sigmaadd}
\Sigma_{cont, i} = \frac{d\theta \,d\phi}{4\,r_{\rm p}^{2}} \Sigma_{\rm n}.
\end{equation}

This expression is formed by considering the mass in the overlapping area given by $d\theta \, d\phi$. If the pixel is totally covered by the node, then it gets the full column density of the node. If the pixel is only partially covered by the node, then the mass in the overlapping region is smeared out over the area of the pixel, to create a new column density. If the node is totally contained within the pixel, then obviously all the mass from the node is smeared out over the pixel's area.

Clearly, the ability of $\theta$ and $\phi$ to describe the overlapping area breaks down near the poles, with the extreme case where a pixel directly over either of the poles cannot be described by a $d\phi$. To account for this, we move to a co-ordinate system in which the tree node's position vector, $\bf n$ describes a new $x$-axis (${\bf x}'$), such that the node is always located at (1, 0, 0). To define the other two axis (the new $y$ and $z$ axis), we first define a control vector, $\bf a$, that is close to the node's position vector, displaced by a small amount in $\theta$ and $\phi$.  For the displacement we choose (somewhat arbitrarily) that $\theta$ decreases by $\frac{1}{2} r_{\rm p}$ for $\theta < \frac{\pi}{2}$ and  increases by $\frac{1}{2}r_{\rm p}$ for $\theta >\frac{\pi}{2}$, and that $\phi$ always increases by $\frac{1}{2}r_{\rm p}$ We then define the new axis vectors from:
\begin{equation}
{\bf x'} = {\bf n}
\end{equation}
\begin{equation}
{\bf b} = {\bf x}' \times {\bf a} 
\end{equation}
\begin{equation}
{\bf y'} = \frac{\bf b}{|\bf a|} 
\end{equation}
\begin{equation}
{\bf z'} = {\bf x}' \times {\bf y}'. 
\end{equation} 
\noindent The pixel unit vectors $\bf p$ are then rotated into this co-ordinate system simply by taking the scalar product with each of the axes:
\begin{equation}
p_x' = {\bf p } \cdot {\bf x}'
\end{equation}
\begin{equation}
p_y' = {\bf p } \cdot {\bf y}'
\end{equation}
\begin{equation}
p_z' = {\bf p } \cdot {\bf z}'.
\end{equation}
\noindent A schematic diagram of how the pixels and nodes are defined in this new coordinate system is given in Figure \ref{fig:nodepro}. The angular distances $R_{\theta}$ and $R_{\phi}$ can then be defined simply from:
\begin{equation}
R_{\theta} = {\rm arcsin}\, p_z'
\end{equation}
\noindent and,
\begin{equation}
R_{\phi} = {\rm arccos}\, \frac{p_x'}{\sqrt{p_x'\,^2 + p_y'\,^2}}.
\end{equation}
\noindent To increase the speed of the algorithm, these inverse trigonometric functions can be made into look-up tables. 

It should be noted that after this rotation, the pixels -- and even the tree node itself -- are in general {\em not} aligned as they appear in Figure \ref{fig:nodepro}. In the above coordinate transform, the control vector $\bf a$ determines how the new coordinate vectors $y'$ and $z'$ are orientated with respect to the new x axis, $x'$ (that is, how $y'$ and $z'$ are rotated {\em around} $x'$). In fact, it should also be stressed that the {\em HEALPix} pixels are not aligned as in Figure \ref{fig:nodepro} {\em before} the rotation, but in fact appear more diamond shaped (as one can see in the maps in Figures 5 - 9).  We found that the exact rotation is typically unimportant for the mapping, provided the angular resolution in the map is not significantly smaller than the opening angle used during the tree-walk. We discuss this further in Section \ref{tests}.

In our discussion so far, we have referred only to `nodes', and their properties, but it should be stressed that some of the nodes will be `leaves'. In our implementation, the leaves are SPH particles, and as was mentioned above, it is customary to use the particle properties directly when evaluating the gravitational forces (or in our case, the column density). As such, we adopt the exact particle position when considering the leaf nodes. However, although SPH particles have non-uniform radial column density profiles, we do not take this into account in our \treecol~implementation, but rather treat the SPH particles in the same manner as the other nodes, by assuming that they have a uniform column density, and project a square in the sky, rather than a circle. As our node to pixel mapping is based around mass conservation (the concept behind Equation \ref{sigmaadd}), we cannot simply use the smoothing length $h$ to define the square (so $r_{\rm n} = h$ in Gadget~2, or $2\,h$ for the definition of $h$ in most other SPH codes), but instead have to define $r_{\rm n}$ to conserve area, giving,
\begin{equation}
r_{\rm n} = \frac{1}{2} \sqrt{\pi} h
\end{equation}
\noindent Our motivation for treating the SPH particles in this way, is that working out the true fraction of the SPH particle's mass that falls within the pixels is computationally expensive, requiring a numerical integration over the overlapping areas, since the pixel and the SPH particle have quite different shapes. 

Our choice of the {\em HEALPix} method of pixelating the spheres around the SPH particles was motivated by two main factors. First, the pixels in the {\em HEALPix} mapping are equal area, which simplifies any comparisons of the pixel properties between different maps. Second, the equal-area property means that increasing the pixel number is equivalent to increasing the angular resolution of the map {\em everywhere on the sphere}. This is not the case with the traditional latitude-longitude discretisation, for example, in which the pixels at the poles have a significantly smaller area than their counterparts at the equator.

Finally, for our SPH code, we use the publicly available code Gadget~2 \citep{springel05}, which uses an oct-tree. For this project, where we need to have control over the opening angle $\theta_{\rm tol}$ to show how it affects the results, we adopt the standard Barnes and Hut opening criterion \citep{bh89} rather than the `relative error' criterion suggested by \citet{springel05}.

\section{Tests of TreeCol}
\label{tests}

In this section we apply the \treecol~algorithm to two very different types of test problem. In the first case, we consider a gas cloud that is an isolated sphere, with conditions similar to the dense cores found in the Pipe nebula \citep{Alves2007}. For the second test problem, we consider a cloud that is a model of a turbulent molecular cloud, which is representative of the environment in which prestellar cores form. These two different set-ups are typical of those used in contemporary simulations of star and cluster formation.

In what follows, we will use Hammer projections to display the $4 \pi$ steradian maps of column densities seen from given locations within the cloud. There are four types of map that we will show. The first is the `true' column density map obtained by summing up the contribution from every single SPH particle in the simulation. For this type of map, it is customary to take into account the radial density profile of the SPH particles,  as described by their smoothing kernel. However since this is not done in the \treecol~implementation we choose not to do this here for the SPH maps. Instead we assume that each SPH particle has a constant column density defined simply by its mass, and radial extent (that is, the particle's smoothing length).
 
The second type of map is the `pixel-averaged' map, whereby we pixelate the `true' map into a set number of {\em HEALPix} pixels, by averaging over the points from the Hammer projection that lie inside each pixel. This type of map provides a more useful measure than the `true' maps, as the results of \treecol ~are also stored on {\em HEALPix} grids. As such, \treecol~could be said to be working perfectly if it can recover the same column densities as those shown in the `pixel-averaged' maps.

The third type of map is simply the column density map produced by \treecol. Finally, our last type of map describes the error in the \treecol~method. 
We define a fractional error $f_{i}$ for each pixel $i$ by
\begin{equation}
f_i  = \frac{|\Sigma_{t_i} - \Sigma_{p_i}|}{\Sigma_{p_i}},  \label{frac_error}
\end{equation}
\noindent where $\Sigma_{p_i}$ is the column density of pixel $i$ in the `pixel-averaged' map, and $\Sigma_{t_i}$ is the column density in pixel $i$ recovered by \treecol. 

In our tests, we will also explore the two intrinsic resolutions that are at play in our implementation of \treecol. The first is the number of pixels in the {\em HEALPix} sphere that surrounds the SPH particles, which represents the ability of the SPH particle to record the column density information that comes from the tree walk. The second resolution at play is the opening angle, $\theta_{\rm tol}$, as this determines how accurately the tree is forced to look at the structure in the cloud. Together these determine the accuracy and level of detail that is present in the \treecol~map.

\begin{figure}
\includegraphics[width=3.4in]{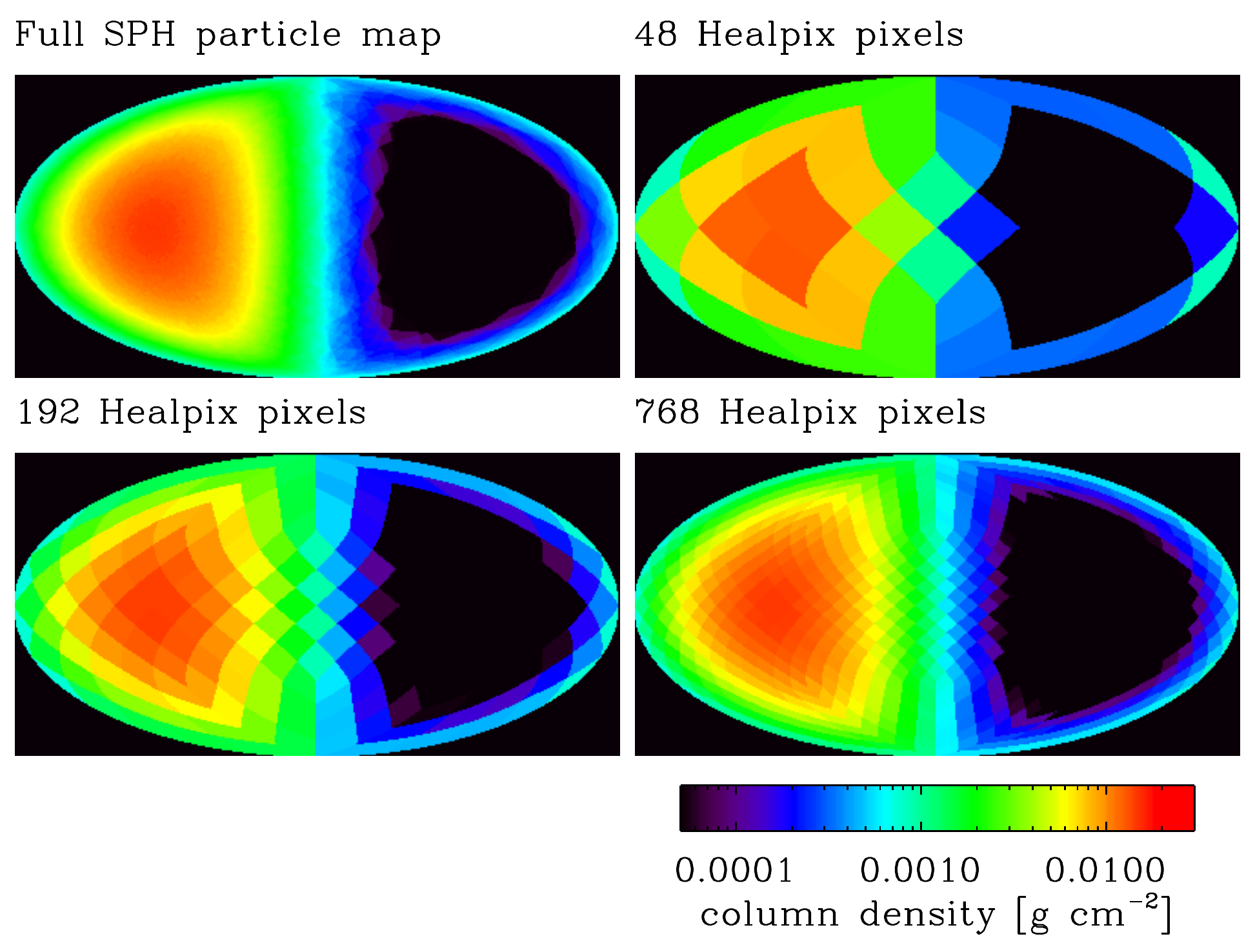}
\caption{\label{spheretrue} The column density Hammer projections for a particle sitting on the edge of a centrally condensed sphere. The upper left panel shows the column density projection from all SPH particles in the simulation volume. The other panels then show the same $4\pi$ steradian map pixelated into 48, 192, and 768 {\em HEALPix} pixels. The pixel values are simply averages of Hammer projection points that lie inside each pixel's boundary.}
\end{figure}

\begin{figure}
\includegraphics[width=3.4in]{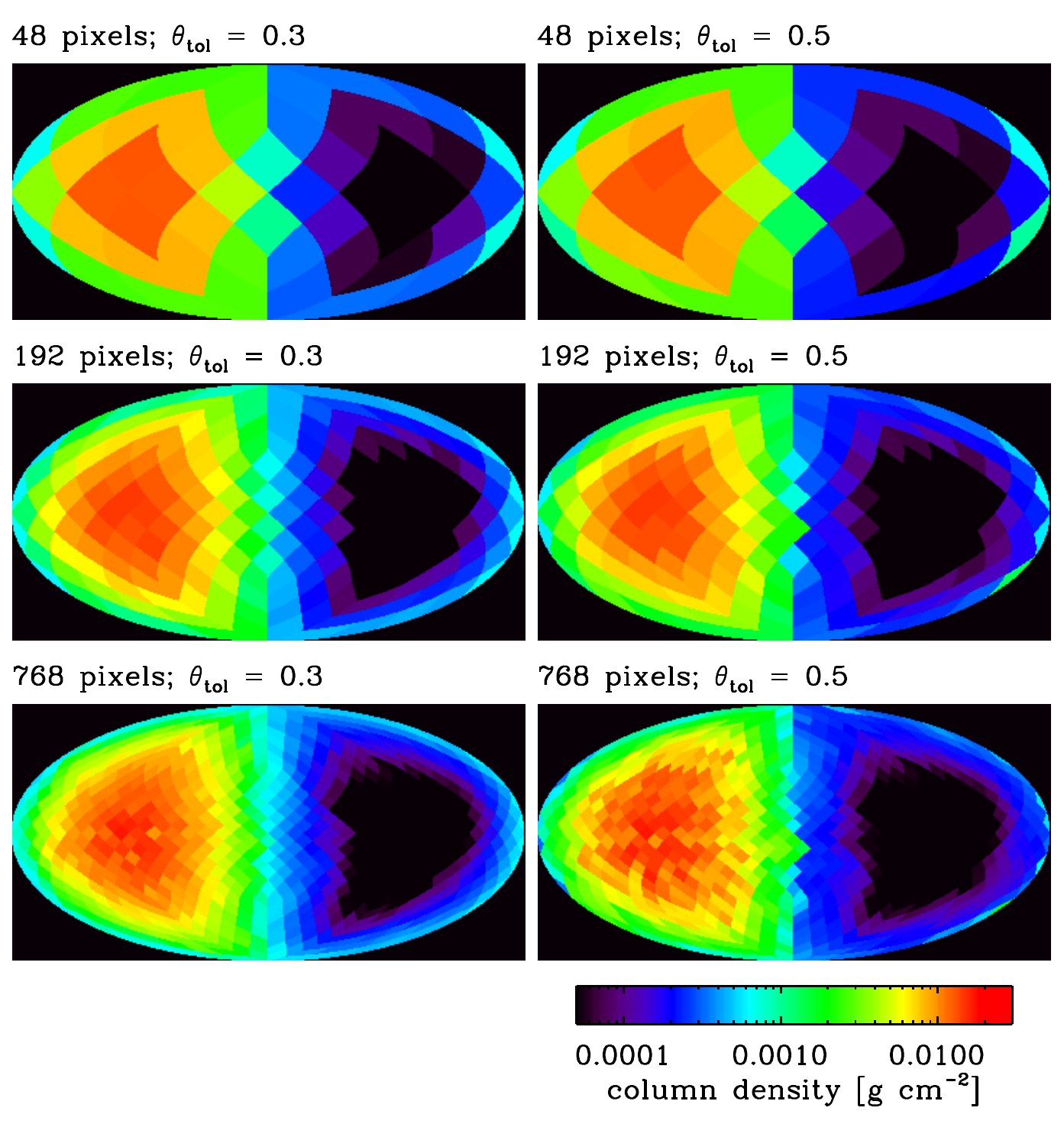}
\caption{\label{spheretreecol} The maps recovered by our \treecol~implementation, for the column density distribution shown in Figure~\ref{spheretrue}. The maps are shown for two different measures of the resolution: the opening angle, $\theta_{\rm tol}$, of the tree (a measure of how well \treecol~can `see' the cloud), and the number of pixels in the {\rm HEALPix} map (a measure of how accurately \treecol's results are stored).}
\end{figure}

\subsection{Spherical cloud}
\label{sec:spheretest}

In the first test problem, we consider a particle located at the edge of a spherical, isothermal cloud with a mean mass density of $3 \times 10^{-20}$ g\,cm$^{-3}$, a temperature of 10~K and a mass of 1.33 \solmasp. The cloud is modelled with 10,000 SPH particles, and hence the mass resolution is comparable to that used in
contemporary models of cluster formation (e.g. \citealt{Bate03, ClarkBonnell05, Jappsen05}). The cloud is gravitationally unbound, but confined by an external pressure of $10^6 \: \rm{K\, cm^{-3}}$ and has been allowed to settle into hydrostatic equilibrium. It centrally condenses into a stable Bonnor-Ebert sphere \citep{Ebert55, Bonnor56, Ebert57} with a central density of $3.4 \times 10^{-20}$ g\,cm$^{-3}$ and an outer density of $1.7 \times 10^{-20}$ g\,cm$^{-3}$ at a radius of 0.09 pc. The column density map of the sky, as seen by the particle at the edge, is shown in Hammer projections in Figure \ref{spheretrue}. In this Figure, we show the `true' column density map, as calculated from the individual SPH particles that make up the cloud, and also the map pixelated into 48, 192, and 768 {\em HEALPix} pixels to create the `pixel-averaged' maps described above. On the left-hand side of the Hammer projections one can see the high column density of the centrally condensed core of the sphere, and on the right-hand side of each map one can see the edge of the sphere, and the empty void beyond.

Although such a simple cloud geometry may seem trivial, it actually represents a stringent test of the \treecol~algorithm. First, the tree itself is made up of a series of boxes, and so the intrinsic geometries of the cloud and tree are quite different. Second, we would expect that the  rapidly evolving gradient in the column densities -- associated with the sharp edge of the cloud -- will be difficult for the tree to capture, as the edges of the nodes will tend to be in a different place, as discussed above.

Despite these difficulties, the algorithm is able to capture the main features of the cloud fairly well. Figure \ref{spheretreecol} shows the \treecol~representation of the sky maps given in Figure \ref{spheretrue} for two different tree opening angles, $\theta_{\rm tol} = 0.3$ and $\theta_{\rm tol} = 0.5$. We can see that the column density towards the centre of the cloud is well represented, and that the maps have the same overall features as those in Figure \ref{spheretrue}: high column density on one side, and a fairly sharp decline on the other side where the column density falls to zero.

\begin{figure}
\includegraphics[width=3.4in]{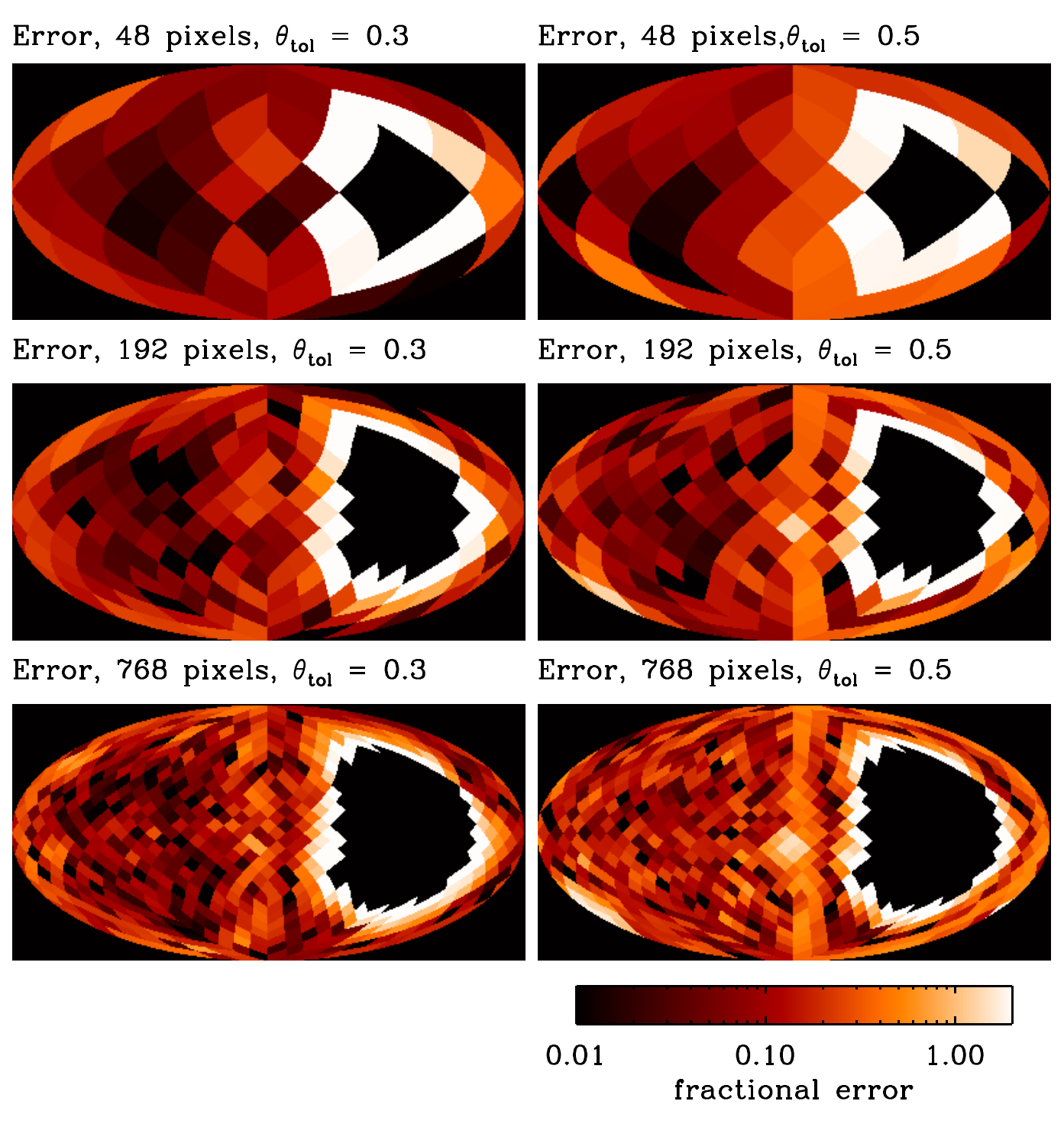}
\caption{\label{sphereerror} The relative error (computed according to Equation \ref{frac_error}) based on the difference between the maps shown in Figure \ref{spheretreecol} and the pixellated maps shown in Figure \ref{spheretrue}.}
\end{figure}

Although the images in Figure \ref{spheretreecol} give an idea of the structure and boundaries that \treecol~is able to reproduce, it is difficult to gauge the quantitative accuracy of the method. A better representation is shown in Figure \ref{sphereerror}, where we plot the relative error in the \treecol~maps. Here we see that the error is typically less than 10 percent when the column density is high, but can be as large as around 100 percent when the column density is low, or approaching zero. The high error (around 50 percent) in the middle of the map (and the outer extremities) comes from the fact that the boundaries of the tree nodes are not necessarily aligned with the edge of the particle distribution. As we increase \treecol's ability to see the structure in the cloud, by reducing the opening angle, we see that the error at the boundary decreases. Overall, the best representation of the cloud's boundary (and indeed the cloud itself) is found in the 48 pixel map that was run with a tree opening angle of 0.3. This is unsurprising, as the low resolution of the pixel-averaged map is also unable to capture the sharp fall in the column density at the cloud's boundary, while at the same time the smaller opening angle ensures that the pixels on the boundary are not assigned mass that belongs to further inside the cloud. 

In general, we see that the smaller opening angles tend to produce better maps for a given pixellation. This is expected, since as the opening angle is reduced, the properties of the tree nodes become closer to the actual distribution of the particles. This is most obviously apparent in the 768 pixel map, where we see that the map obtained for $\theta_{\rm tol} = 0.5$ contains artefacts from the underlying boxy structure of the tree, while the $\theta_{\rm tol} = 0.3$ map is much smoother. In the maps with a 
lower number of pixels, these features are not so apparent as the structure of tree is more smeared out. 

Perhaps a more useful measure of the ability of \treecol~to sample its surroundings is the error in the average column density in the map, as given in Table \ref{meancol}. For the spherical cloud set-up, we find that the average column is between 4.2 and 7 percent higher than the average in the true map, with the lowest resolution run ($\theta_{\rm tol} = 0.5$, N$_{\rm pix}$ = 48) having the largest overall error, and the highest resolution run ($\theta_{\rm tol} = 0.3$, N$_{\rm pix}$ = 768) having the lowest error. The fact that these errors are so low reflects the fact that the mean is dominated by the high column density regions, which are recovered well by \treecol~in all the resolutions we study.

\begin{figure}
\includegraphics[width=3.4in]{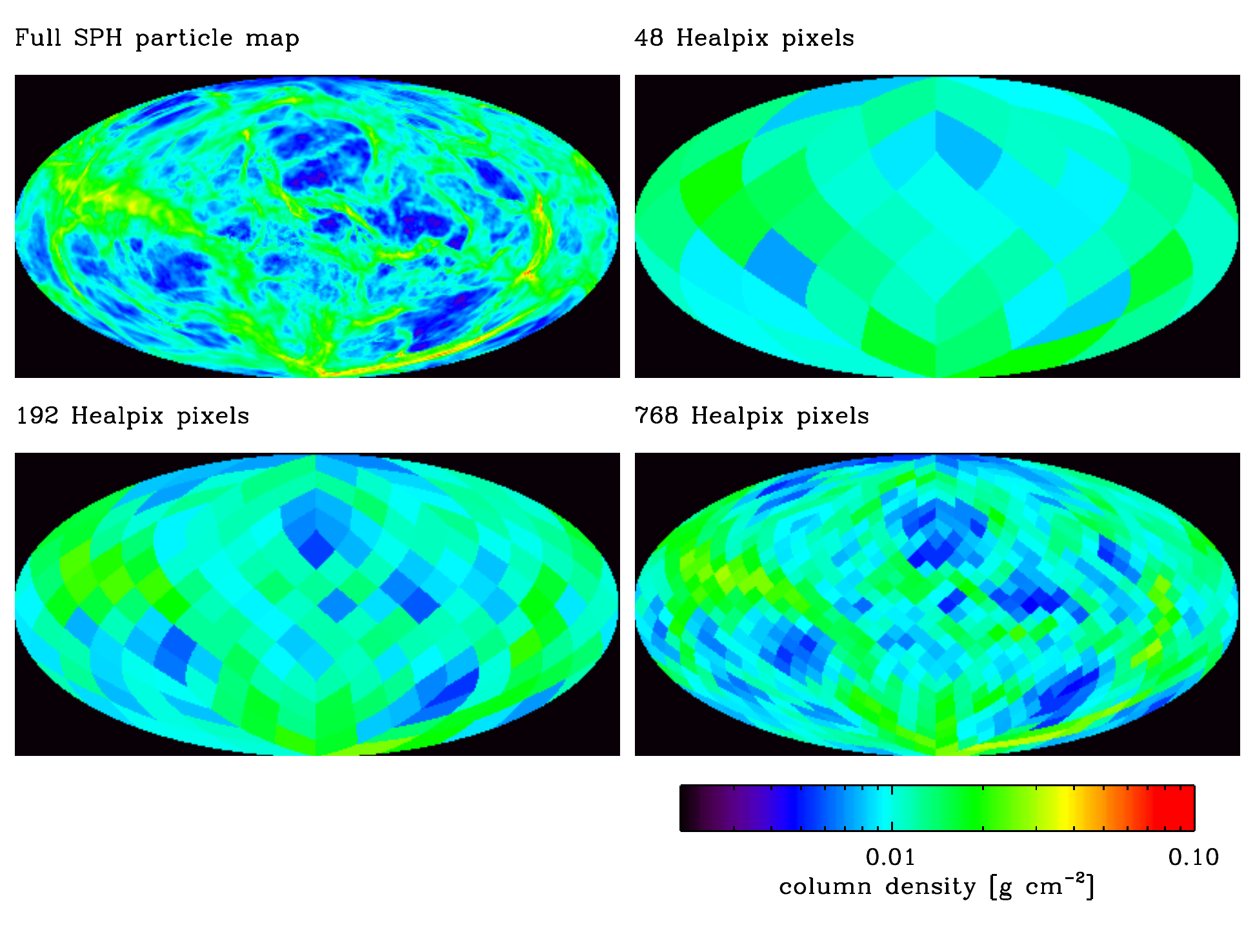}
\caption{\label{turbtrue} The column density sky-map as seen by a low-density particle in a turbulent molecular cloud simulation. As in Figure \ref{spheretrue}, the upper-left panel is obtained by adding up the contributions from all SPH particles in the computational volume (excluding the particle from which the sky is viewed). The other panels then show a `pixel-averaged' view of the cloud, as would be seen if we only had 48, 192 and 768 pixels in our map.}
\end{figure}

\begin{figure}
\includegraphics[width=3.4in]{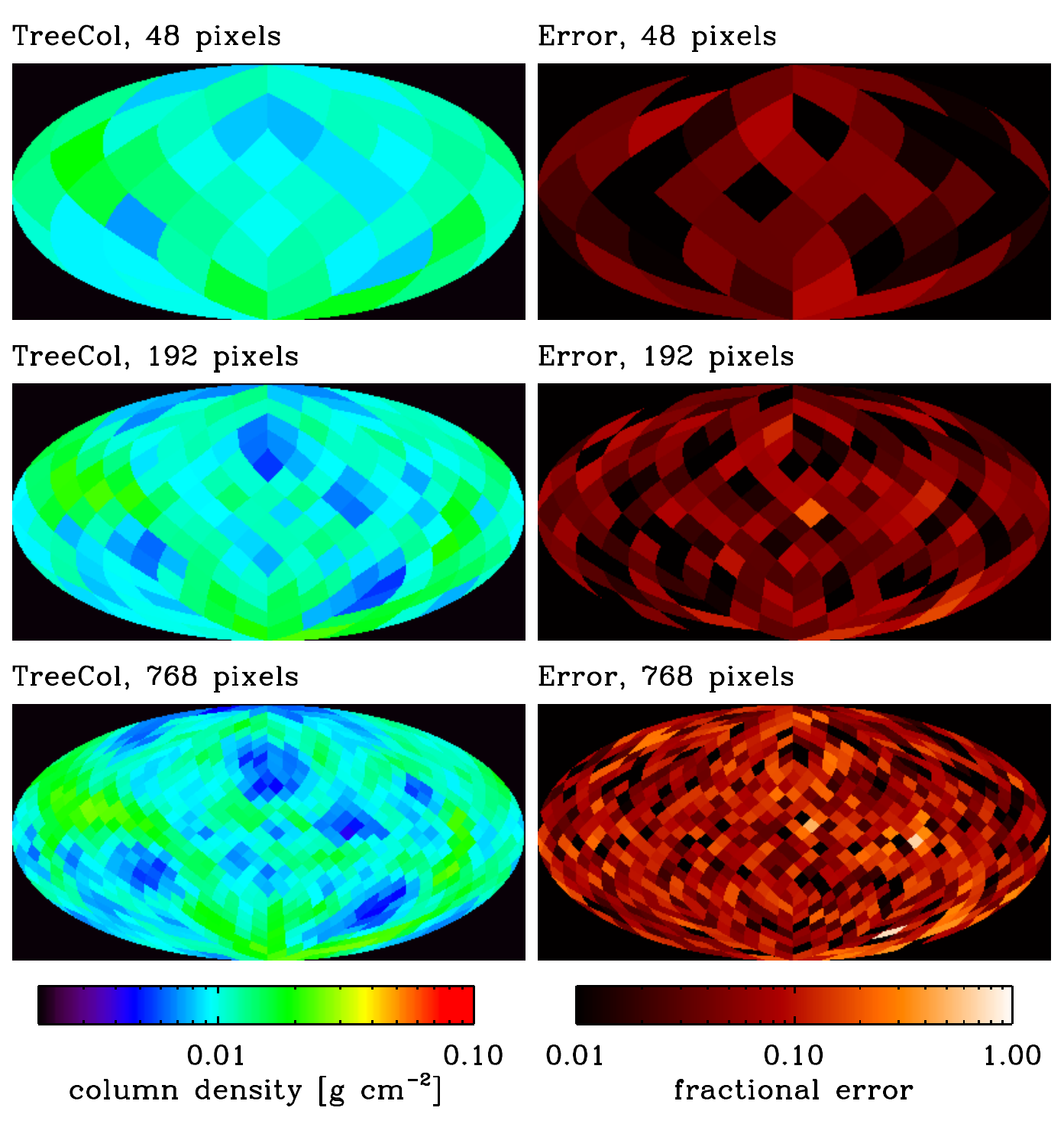}
\caption{\label{turbtreecol} The left-hand panels show the \treecol~maps for the turbulent cloud set-up shown in Figure \ref{turbtrue}, for 48, 192 and 768 pixels in the map. All maps were produced using a tree opening angle $\theta_{\rm tol} = 0.3$. The right-hand panels show the relative error in the \treecol~maps.}
\end{figure}

\begin{table}
\caption{A summary of the mean column densities in the cloud models presented in Sections \ref{sec:spheretest} and \ref{sec:turbtest}, for both the true map (the first line) and each of the \treecol~maps. For the \treecol~results we give the number of pixels used in the column density map (N$_{\rm pix}$), the opening angle of the tree ($\theta_{\rm tol}$), and the percentage error compared to the true map from the SPH particles. Note that due to the way the pixel-averaged maps are obtained (see Section \ref{tests}), their average column density is identical to that in the full SPH map, and so we do not include it here.}
\begin{tabular}{ c | c | c | c | c}
Model  &  N$_{\rm pix}$  &  $\theta_{\rm tol}$  &  $\bar{\Sigma}$   &  Error    \\
           &                           &                                    & $[\rm g\,cm^{-2}]$ & [\%] \\
\hline
\hline
Spherical cloud & & & 3.060 $\times 10^{-3}$ & \\
\hline
	&  48   &  0.3  &  3.234 $\times 10^{-3}$  &  5.7 \\
	&  48   &  0.5  &  3.274 $\times 10^{-3}$  &  7.0  \\
	&  192 &  0.3  &  3.205 $\times 10^{-3}$  &  4.7  \\
	&  192 &  0.5  &  3.239 $\times 10^{-3}$  &  5.8  \\
	&  768 &  0.3  &  3.192 $\times 10^{-3}$  &  4.3  \\
	&  768 &  0.5  &  3.226 $\times 10^{-3}$  &  5.4  \\
\hline \hline
Turbulent cloud & & & 1.151 $\times 10^{-2}$ & \\
\hline
	&  48   &  0.3  &  1.126 $\times 10^{-2}$ &  2.2 \\
	&  192 &  0.3  &  1.125 $\times 10^{-2}$ &  2.3 \\
	&  768 &  0.3  &  1.133 $\times 10^{-2}$ &  1.6  \\
\hline
\end{tabular}
\label{meancol}
\end{table}

\subsection{Turbulent clouds}
\label{sec:turbtest}

Our previous test examined the ability of the \treecol~algorithm to capture the column density variations that one would expect in the environment of a prestellar core. However the test was also designed to see how well \treecol~can handle sharp density contrasts, and so the core was simply placed in a vacuum, rather than the more complicated environment of a turbulent molecular cloud, in which the typical prestellar core is born \citep{MacLowKlessen2004}. This is the focus of the test in this section. Again, we want to make the test problem as challenging as possible, so in this section we choose to examine the sky-map as seen by a low density particle in the cloud. For such a particle, the holes and filaments that characterise the turbulent cloud's structure should be more pronounced than they would be for a high density particle, and so the contrast in the column density map is high.

Our cloud has a mass of 10$^{4} \: {\rm M_{\odot}}$, an initial mean number density of 300 cm$^{-3}$ (or mean mass density $1.17 \times 10^{-21}$ g\, cm$^{-3}$), and a radius of roughly 6 pc. We model the cloud with $2 \times 10^6$ SPH particles. At the start of the simulation, we impose a turbulent velocity field on the cloud that has a power spectrum of the form $P(k) \propto k^{-4}$, and adjust the strength of the velocities such that the kinetic energy in the cloud is equal to the gravitational energy of the cloud. This gives an initial root-mean-squared velocity of around 3 km\,s$^{-1}$. The turbulence is left to freely decay in shocks as it creates structure in the initially uniform density cloud.

We stop the calculation after a period of $6.4 \times 10^5$ yr, by which time a combination of the turbulence and self-gravity has created a network of interconnected filaments and voids. This structure can been seen in the sky-maps shown in Figure \ref{turbtrue}, where, as discussed, we position ourselves on a low-density particle that resides near the centre of the cloud. As in Figure \ref{spheretrue}, we again show how this map would look if it were to be de-resolved to 48, 192, and 768 HEALPix pixels, giving us some idea how well \treecol~can be expected to perform. It is already obvious from the pixel-averaged maps that even at the 768 pixel  level, many of the very dense features are going to be missing from the map. Nevertheless, the mean column densities in the coarse, pixellated maps are all within 0.1 percent of the mean column in the full SPH map, and so they are still a good representation of the column density distribution in the cloud, even if they are unable to resolve the small-scale detail.

The images in Figure \ref{turbtreecol} show results from \treecol~for this cloud, including the \treecol~ column density maps and their associated relative errors. Given the amount of structure in the cloud, we construct the maps in this figure while keeping the tree-opening angle fixed at 0.3. Overall we see that the algorithm behaves well, and the features present in the pixel-averaged maps are recovered, even at our highest mapping-resolution of 768 pixels. For the 2 lower-resolution maps (48, and 192 pixels), the errors in the maps are mainly small, and \treecol~ typically recovers the column densities to around 5 percent. However, we see that the errors in the 768 pixel map are again quite high,  and for the same reasons as we seen in the previous test, namely that the pixellation of the map is too high for the adopted tree-opening angle, and so the structure of the tree is beginning to show in the map.

Although the cloud studied here is more complicated than that studied in Section \ref{sec:spheretest}, the errors in the mean column density (given in Table \ref{meancol}) are actually lower than they are in the spherical cloud, and range from 1.6 to 2.3 percent. Unfortunately, the extra small-scale structure means that the way in which the error relates to the number of pixels is not as consistent here as it was for the previous cloud set-up. As one moves to higher number of pixels, the small-scale, high-column features start to become resolved, but although \treecol~is able to see them, it tends to get their location wrong as the tree node that represents these regions occupies a slightly different part of the sky, and has a different angular extent, than the true SPH particle distribution. So while the mass located in these small scale features is captured by the map, the location and spread is not, and as a result, one pixel may get too much column, while a neighbour receives too little. This is why the pixel error (remember that this is the absolute error) in Figure \ref{turbtreecol} starts to get worse as the number of pixels increases.

\subsection{Dust heated by the interstellar radiation field}

\begin{figure}
\includegraphics[width=3.4in]{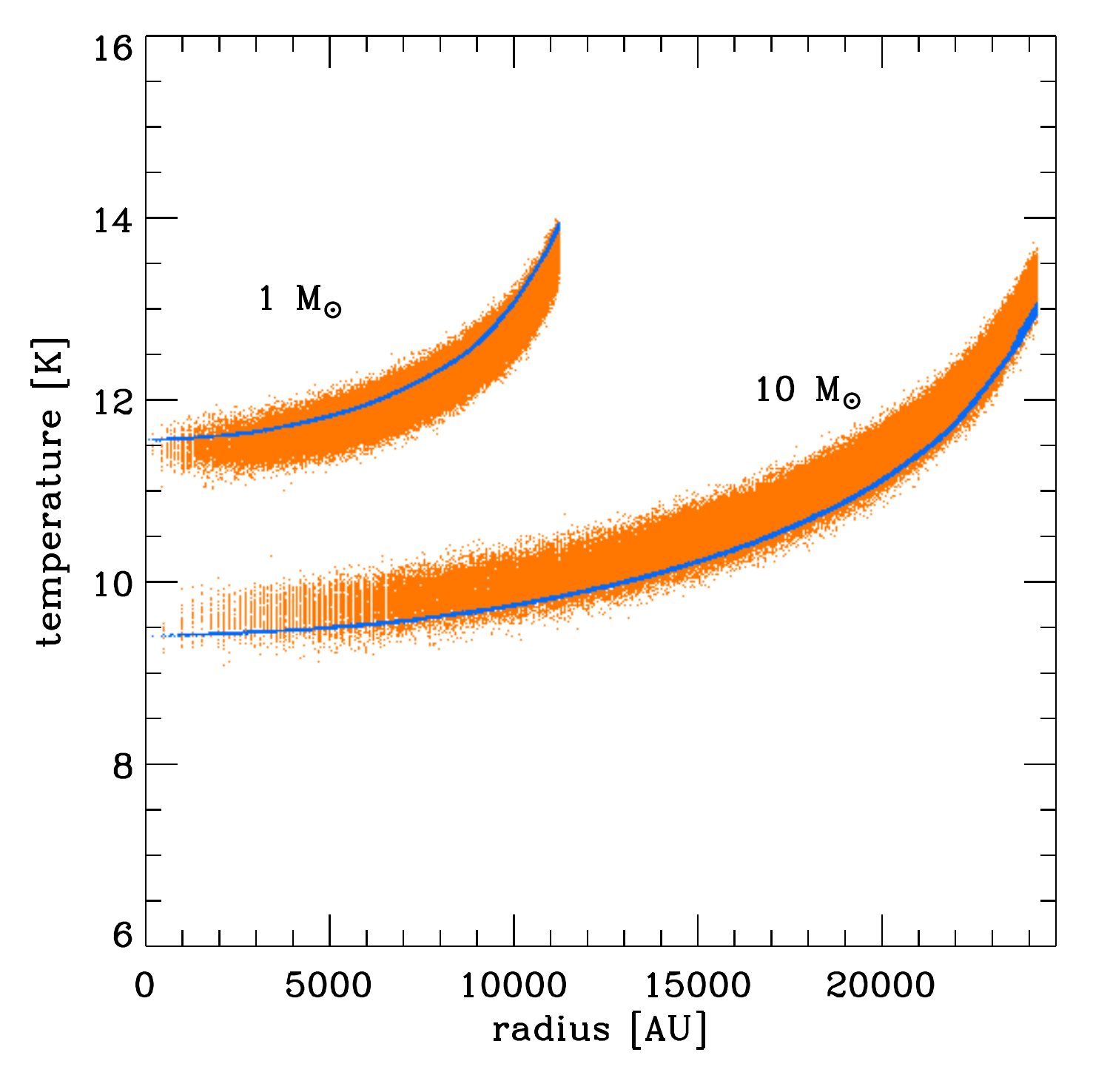}
\caption{\label{tdust_profiles} The dust temperature profiles for two uniform density clouds ($10^{-19} \rm g\,cm^{-3}$) of mass 1 and 10 \solmasp, heated by the Black (1994) interstellar radiation field. Orange points show the output from the RADMC-3D Monte Carlo radiative transfer code -- run with $80^3$ grid cells and $2 \times 10^7$ photon packets -- and the blue points denote the output from an SPH simulation that uses the column density information recovered by \treecol~in conjunction with the method for calculating dust temperatures given in Goldsmith (2001). In the SPH simulation we use 261932 particles and a tree-opening angle of 0.5. The dust opacities are a combination of \citet{oh94} (non-coagulated and thick ice mantle grains) for wavelengths longer than $1 \: \mu{\rm m}$, and those given in \citet{mmp83} for shorter wavelengths.}
\end{figure}

Although we have seen that \treecol~can typically deliver a fairly accurate column density map of the sky, there are situations in which the errors in the map can be as much as 100 percent, and so it is prudent to check whether this is a problem when one applies \treecol~to a real astrophysical calculation. To this end, we look at a typical problem in star formation: the temperature profile of prestellar cores, heated by the interstellar radiation field (ISRF). This problem has been looked at by number of authors, with aims ranging from deriving observational masses from dust emission \citep[see e.g.][]{stam07}, to understanding the dynamics and fate of prestellar cores \citep{kc08,kc10}. Furthermore, Larson (2005) suggested that the dust temperature at the onset of thermal coupling between the gas and dust could help set a characteristic Jeans mass in molecular clouds, which may be responsible for setting the characteristic mass for star formation (see also \citealt{Jappsen05}). Since calculating the dust temperatures reduces to evaluating the attenuation of the ISRF -- which depends on the column density between the core and the incoming radiation -- it is ideally suited to the \treecol~approach. We describe here a simple method by which this can be included into our \treecol~implementation.

In the case of a static cloud in thermal equilibrium, one can solve for the dust temperature $T_{\rm d}$ by finding the value that satisfies the equation of thermal balance for the dust:\footnote{Note that we assume here for simplicity that energy transfer from the gas to the dust (or vice versa) does not significantly affect the dust temperature. 
This assumption is valid whenever that the gas and dust temperatures are equal or the gas density is low enough that the coupling between dust and gas is weak.
For a more detailed treatment, see e.g.\ \citet{gc2011a}.}
\begin{equation}
\label{dust_temp}
\Gamma_{\rm ext} - \Lambda_{\rm dust} = 0.
\end{equation}
Here $\Gamma_{\rm ext}$ is the dust heating rate per unit volume due to absorption of radiation from the ISRF and $\Lambda_{\rm dust}$ is the radiative cooling rate of the dust.

Following \citet{gold01}, one can express $\Gamma_{\rm ext}$ as the product of a optically thin heating rate, $\Gamma_{\rm ext, 0}$, and a dimensionless factor, $\chi$, that represents the attenuation of the ISRF by dust absorption:
\begin{equation}
\label{dust_heat}
\Gamma_{\rm ext} = \chi \Gamma_{\rm ext, 0}.
\end{equation}
The optically thin heating rate is given by
\begin{equation}
\Gamma_{\rm ext, 0} = 4\pi {\cal D} \rho \int_{0}^{\infty} J_{\nu} \kappa_{\rm \nu}  \, {\rm d}\nu,
\end{equation}
where ${\cal D}$ is the dust-to-gas ratio, $\rho$ is the gas density, $J_{\nu}$ is
the mean specific intensity of the incident ISRF, 
and $\kappa_{\nu}$ is the dust opacity in units of ${\rm cm^{2}} \: {\rm g^{-1}}$.  Provided the properties of the radiation field and the dust do not change over the region in question, the integral in the above expression needs to be computed only at the start of the simulations, and then simply used as a pre-factor. In our application, we adopt the ISRF given in \citet{bl94}, and the dust opacities of \citet{oh94} (non-coagulated and thick ice mantle grains) for wavelengths longer than $1 \: \mu{\rm m}$, and those from \citet{mmp83} at shorter wavelengths. 

The attenuation factor $\chi$, is then given by the equation
\begin{equation}
\label{heat_chi}
\chi(N_{\rm H}) = \frac{4\pi \int_{0}^{\infty} J_{\nu} \kappa_{\nu} \exp \left(-\kappa_{\nu} \Sigma \right) 
\, {\rm d}\nu}{4\pi \int_{0}^{\infty} J_{\nu} \kappa_{\nu} \, {\rm d}\nu}, \label{chi_func}
\end{equation}
where $\Sigma = 1.4 m_{\rm p} N_{\rm H}$, $m_{\rm p}$ is the proton mass, and $N_{\rm H}$ is the number density of hydrogen nuclei. This equation can also be solved for a wide range of different values of $N_{\rm H}$, and results can then be stored in a look-up table to be called during the simulation.

For the dust cooling rate $\Lambda_{\rm dust}$, one has to solve,
\begin{equation}
\Lambda_{\rm dust}(T_{\rm d}) =  4 \pi {\cal D} \rho \int_{0}^{\infty} B_{\nu}(T_{\rm d}) \kappa_{\nu} 
\, {\rm d}\nu,
\end{equation}
where $B_{\nu}(T_{\rm d})$ is the Planck function for a temperature $T_{\rm d}$. For our choice of dust opacities, we find that the resulting cooling rate is well fit by the function
\begin{equation}
\label{thin_heat}
\Lambda_{\rm dust}(T_{\rm d}) = 4.68 \times 10^{-31} T_{\rm d}^{6}\, n \: {\rm erg} \: {\rm s^{-1}} \: 
{\rm cm^{-3}}
\end{equation}
for dust temperatures $5 < T_{\rm d} < 100 \: {\rm K}$ \citep{gc2011a}.

Using this framework, it is then straightforward to solve for the dust temperature of each SPH particle. First, a value of $\chi$ can be calculated for each \treecol~pixel, based on its column density. Provided the pixels have an equal area (as is the case with the HEALPix scheme employed here), the arithmetic mean of these $\chi$ values can then be used in Equation \ref{dust_heat} to compute the value of $\Gamma_{\rm ext}$ for the particle.  This is then used in Equation \ref{dust_temp}, along with the cooling rate in Equation \ref{thin_heat}, to compute the dust temperature associated with the SPH particle.

An example of this technique, applied to two clouds, is shown in Figure \ref{tdust_profiles}.
Given that the errors in the spherical cloud test in Section \ref{sec:spheretest} were higher than those for the turbulent cloud in Section \ref{sec:turbtest}, we again choose a spherical, isolated cloud as our test bed, so that we can check whether these errors are important in a typical astrophysical set-up. The clouds both have a uniform density of $10^{-19} \rm g\,cm^{-3}$, but differ in mass, with one having a mass of 1 \solmasp, and the other 10 \solmasp. Figure \ref{tdust_profiles} shows the resulting radial temperature profiles for the clouds as computed using \treecol~in conjunction with the method described above, for 261932 SPH particles. This number of particles is comparable to the number used in simulations of the collapse of prestellar cores \citep{Bate1998, ClarkBonnell05, Bate2010}. We adopt an opening angle of 0.5 in this test, and we use 48 pixels in the column density map. For comparison we also show the results from RADMC-3D\footnote{http://www.ita.uni-heidelberg.de/$\sim$dullemond/software/radmc-3d/}, a Monte Carlo radiative transfer code, performed using 20 million photon packets on a $80^3$ uniform grid. This number of grid cells is chosen to ensure that the number of cells {\em inside} the sphere is the roughly the same as the number of SPH particles. Note that scattering is switched off in the RADMC-3D run, and neither is it taken into account in the $\chi$ factor used in the \treecol~implementation. Finally, for ${\cal D}$ we take the standard value for solar metallicity gas.
 
Comparing the results from the \treecol~method to those from RADMC-3D, we see that the former recovers the general properties of the dust temperature profile very well. The temperatures are higher on the outskirts of the cloud, where the gas is exposed to more radiation, and cooler in the centre, where the gas is better shielded. Also, the lower mass (and therefore overall lower mean column density) cloud is hotter than the higher mass cloud, as every part of it sees more of the ambient radiation field. 

Most importantly however, we see that the results from our \treecol-based dust temperature calculation lie within the scatter of the Monte Carlo code, which itself is only around 0.5K. However note that while the error in the RADMC-3D results will get better as more photon packets are used to model the radiation field, we find that the \treecol~method does not get significantly more accurate as the either the opening angle is decreased, or the resolution in the pixel map is increased. The source of the discrepancy is that RADMC-3D is able to treat the absorption, re-emission and then re-absorption of the incoming photons, while the \treecol~method only models the initial absorption. In other words, what we doing in the \treecol~implementation is only an approximation to the full radiative transfer problem. As such, the main source of error in the \treecol~results shown in Figure \ref{tdust_profiles} is not the error in the column density maps obtained during the tree-walk, but the approximations made in using these column densities to calculate the dust temperature. 

\section{Potential applications}
\label{applications}
There are a number of different areas of computational astrophysics in which we expect the \treecol~algorithm to be useful. One important example is the case examined at in the previous section: modelling the penetration of the interstellar radiation field into a prestellar core, and the resultant heating of the dust. At high gas densities, the gas and dust temperatures within prestellar cores are closely coupled, and the dust plays a crucial role in regulating the thermal behaviour of the gas. An accurate determination of how the equilibrium dust temperature changes during the dynamical evolution of a prestellar core is therefore of great importance in studies of the stability of such cores \citep[see e.g.][]{kc10} and so a method to compute this efficiently within a high-resolution three-dimensional simulation is clearly of great utility.

Another example of the kind of problem for which we expect \treecol~to be a valuable approach is the attenuation of ultraviolet radiation within simulated molecular clouds. UV radiation plays a central role in regulating much of the astrochemistry within molecular clouds \citep[see e.g.][]{ht99}, and yet current numerical models of cloud formation typically use very simplistic methods to model the radiation field. For instance, the studies by \cite{dbp06} and \cite{dobbs08} use an approximation in which the absorbing column at  any point in the simulated interstellar medium is estimated by multiplying the local gas density by a fixed shielding length. \citet{gtk09} use a similar local approximation, but rather than keeping the shielding length fixed, determine it by examining the local density and velocity gradients (see also \citealt{gk11}). Finally, the detailed models presented in \citet{glo10} use a ``six-ray'' approach in which the column density of gas between each cell and the boundaries of the simulation is calculated along six lines of sight, taken to run parallel to the coordinate axes. \treecol~represents a significant improvement in accuracy over all of these approaches, and we have already begun to make use of it in our work on star formation within molecular clouds \citep{gc2011b}.

A further example of a possible application for our method is modelling of the X-ray heating of the intergalactic medium (IGM) prior to the epoch of reionization. At high redshifts, X-rays produced by sources such as massive X-ray binaries 
\citep[e.g.][]{gb03,mira11} or mini-quasars \citep{zaroubi07} heat the neutral IGM, producing distinctive signatures in the 21-cm background  \citep[see e.g.][]{pf07}. As the heating is dominated by soft X-rays, with relative short mean free paths, the resulting temperature distribution of the gas is inhomogeneous. Accurate modelling of the temperature distribution is necessary if we are to make optimal use of the information provided by the 21-cm background, and this in turn requires us to compute the column density distribution around numerous sources in a computationally efficient manner, an ideal application for an approach such as \treecol.

Of course, we anticipate that \treecol~will be useful in areas besides those listed here, but hope that this brief discussion gives a general idea of the situations in which an efficient method for estimating column densities within numerical simulations is likely to be useful.

\section{A note on performance}
\label{performance}

The fact that \treecol~makes use of the pre-existing gravitational tree has several advantages when it comes to the performance and implementation of the algorithm. First, it means that the algorithm will scale with increasing particle number in the same way as the tree scales. As discussed above, this is typically $N$ log $N$ for most tree codes (note however that some techniques may make it possible to achieve even better scaling, for example the features discussed in \citealt{gr11}). In contrast, ray-tracing -- the method most commonly used for obtaining column density estimates -- typically scales as $N^{5/3}$. Further, the implementation of \treecol~is relatively easy, as the three quantities that are required to capture a node's column density contribution (namely the mass, relative position and angular size) are already used in the calculation of the gravitational forces. As such, implementing \treecol~requires few (if any) structural changes to the underlying tree code, and mainly reduces to adding a call to a function that handles the mapping of the node's contribution to the target particle's column density map. As such, \treecol~can also be implemented easily in grid-based fluid codes that use a tree-scheme to calculate gravitational forces. For example, the method has recently been implemented into FLASH (A\~{n}orve, private communication).

Another useful feature of such a tree-based algorithm is that it is naturally adaptive: every particle in the cloud will see its immediate surroundings at a set angular resolution, regardless of the physical size of the density structure at the local scale of the particle. As an example, one can take the case of a protostellar disc sitting in a collapsing protostellar core. In \treecol, the particles on the edge of the disc will be able to pick up the local drop in column density arising from the disc-envelope boundary, while those on the edge of the core will pick up the boundary between the core and the ambient cloud. This is a natural consequence of the way that a gravitational tree solver breaks up the cloud into a hierarchically nested grid. 

Also, it should be noted that tree codes tend to parallelize well, and parallel tree gravity is now a standard feature of contemporary SPH codes. As such, \treecol~can naturally take advantage of the speed-up that parallelization offers. In order to verify that the additional work that must be done during the tree-walk need not adversely affect the parallel scaling of the code, we have studied the scaling of a representative test problem both with and without the use of \treecol. For our test, we modelled the evolution of an isothermal, turbulent molecular cloud using the Gadget 2 SPH code, with two million SPH particles. We investigated how the wallclock time required to model the cloud for a specified period varied as we increased $N_{\rm p}$, the number of processors used to run the code. For each value of $N_{\rm p}$, we performed two simulations: one in which \treecol~was used to compute the column densities, and one in which it was not. We defined a parallel efficiency for each simulation as
\begin{equation}
\eta = \left(\frac{N_{\rm p}}{64}\right)^{-1} \frac{T}{T_{64}}
\end{equation}
where $T$ is the elapsed wallclock time for the simulation, and $T_{64}$ is the wallclock time for the $N_{\rm p} = 64$ simulation. (Note that by this definition, $\eta = 1$ for the $N_{\rm p} = 64$ runs). We performed simulations with $N_{\rm p} = 64, 128, 256$ and 512. The results are summarized in Table~\ref{scaling}. 

\begin{table}
\caption{Parallel efficiency of the test calculation described in the text,
performed both with and without \treecol, as a function of the number of 
processors, $N_{\rm p}$. The efficiency is normalized to unity for 
$N_{\rm p} = 64$.
}
\begin{tabular}{ccc}
$N_{\rm p}$ & With treecol & Without treecol \\
\hline
64  & 1.00 & 1.00 \\
128 & 0.61 & 0.61 \\
256 & 0.42 & 0.39 \\
512 & 0.22 & 0.28 \\
\hline
\end{tabular}
\label{scaling}
\end{table}

Although the parallel efficiency of Gadget 2 for this problem is not particularly high to begin with, it is clear that the use of \treecol~does not have very much influence on the scaling, suggesting that the additional communications overhead is not a significant problem in comparison to the inherent difficulties involved in properly load-balancing a simulation of this type.

However, as with any computational method, there are drawbacks to our approach. One of the main downsides of the \treecol~method is that it can introduce a significant memory overhead. The exact memory requirements of \treecol~can vary considerably, depending on the type of tree employed by the code, how the column density information is being used, and whether the code is parallelized using the Message Passing Interface (MPI) protocol, or using the OpenMP protocol. In Gadget~2 for example -- a code that is MPI parallelized -- copies of the SPH particles on a given CPU are sent to all the other CPUs to get the contributions to the gravitational force from the particles that reside there. An implementation in Gadget~2 must then store two copies of the column density map for each particle: one that is broadcast to the other CPUs to pick up their contributions, and one that resides on the home CPU that collects the local contributions and stores the final total. Other tree codes parallelized
using MPI work differently, sending the necessary information {\em from} the other CPUs to the CPU with the target particle, requiring that only one map be stored per particle. In fact, if the column density map is only needed once -- for example, to compute a mean extinction -- then only the particle currently walking the tree needs a column density map. In this case the information stored in the map can be used at the end of particle's walk, and the map can then be cleared in preparation to be re-used in the next particle's tree-walk. So depending on the application, and on the code, the memory requirements for \treecol~can be anywhere from one map per parallel task, to two maps per particle.

\section{Summary}
\label{summary}

We have present a new tree-based technique for obtaining a full $4 \pi$ steradian map of the column densities at every location in numerical fluid simulations. The method piggy-backs on a tree-based gravitational force calculation, by making use of the information that is already stored in the tree -- namely the mass, position, and size of the tree nodes --  to construct a map of the column density distribution in the sky as seen by each fluid element. As the underlying algorithm is based on the tree, the method inherits the same $N\,\log \,N$ scaling as the tree code. The fact that the method makes use of physical quantities that are already stored in the tree means that it is simple to implement, and requires only minimal modification to the underlying tree algorithm. In the case where the tree has been parallelised, we find that the inclusion of \treecol~ does not significantly affect the parallel scaling of the code.

In this paper, we describe a simple implementation of \treecol~that we find to yield column density maps that are accurate to better than 10 percent on average. In this implementation -- which in our case was made within the publicly available SPH code Gadget 2 \citep{springel05} -- we adopt the {\em HEALPix} \citep{healpix} pixelisation scheme to define the pixellated map in which the column densities for each particle are stored. 

As an example application of \treecol~we show how the method can be used to calculate the dust heating of prestellar cores by the interstellar radiation field. The results are compared with those from the Monte Carlo radiative transfer code RADMC-3D. Comparing our lowest resolution \treecol~results -- 48 pixels in the $4 \pi$ steradian {\em HEALPix} map and a tree opening angle of 0.5 -- to a 20 million photon packet RADMC-3D calculation, we find that the two methods yield radial dust temperature profiles that agree to within 0.5K. We also discuss some other applications in which we expect \treecol~to be useful, such as the attenuation of UV radiation and its effect on the chemical and thermal balance of molecular clouds or the X-ray heating of the intergalactic medium.

\section{Acknowledgements}

We would like to thank Mordecai-Mark Mac Low, Tom Abel, Gabriel A\~{n}orve, and L\'{a}szl\'{o} Sz\H{u}cs, for many interesting discussions regarding the \treecol~method, and help with assembling the final manuscript. The authors acknowledge financial support from the Landesstiftung Baden-W\"urrtemberg via their program Internationale Spitzenforschung II (grant P-LS-SPII/18), from the German Bundesministerium f\"ur Bildung und Forschung via the ASTRONET project STAR FORMAT (grant 05A09VHA), from the DFG under grants no.\  KL1358/10 and KL1358/11, and via the SFB 881 ``The Milky Way Galaxy'', as well as from a Frontier grant of Heidelberg University sponsored by the German Excellence Initiative. The simulations reported on in this paper were primarily performed using the {\em Kolob} cluster at the University of Heidelberg, which is funded in part by the DFG via Emmy-Noether grant BA 3706.

%
%

\end{document}